\documentclass[prl,twocolumn,showpacs,amsmath,amssymb,superscriptaddress]{revtex4-1}
\usepackage{graphicx,bm,color,mathptmx,hyperref} 
\newcommand{\Tr}{\mathop{\mathrm{Tr}}\nolimits}
\newcommand{\op}[1]{\hat{#1}}

\begin{document}

\title{Time-Multiplexed Measurements of Nonclassical Light at Telecom Wavelengths}

\author{G.~Harder}
\affiliation{Department of Physics, University of Paderborn, 
Warburger Stra{\ss}e 100, 33098 Paderborn, Germany}

\author{C.~Silberhorn}
\affiliation{Department of Physics, University of Paderborn, 
Warburger Stra{\ss}e 100, 33098 Paderborn, Germany}
\affiliation{Max-Planck-Institut f\"ur die Physik des Lichts,
  G\"{u}nther-Scharowsky-Stra{\ss}e 1, Bau 24, 91058 Erlangen,
  Germany}

\author{J.~Rehacek}
\affiliation{Department of Optics, Palacky University, 
17. listopadu 12, 77146 Olomouc, Czech Republic}

\author{Z.~Hradil}
\affiliation{Department of Optics, Palacky University, 
17. listopadu 12, 77146 Olomouc, Czech Republic}

\author{L.~Motka}
\affiliation{Department of Optics, Palacky University, 
17. listopadu 12, 77146 Olomouc, Czech Republic}

\author{B.~Stoklasa}
\affiliation{Department of Optics, Palacky University, 
17. listopadu 12, 77146 Olomouc, Czech Republic}

\author{L.~L.~S\'{a}nchez-Soto}
\affiliation{Departamento de \'Optica, Facultad de F\'{\i}sica, 
Universidad Complutense, 28040 Madrid, Spain}
\affiliation{Max-Planck-Institut f\"ur die Physik des Lichts,
  G\"{u}nther-Scharowsky-Stra{\ss}e 1, Bau 24, 91058 Erlangen,
  Germany}

\begin{abstract}
  We report the experimental reconstruction of the statistical
  properties of an ultrafast pulsed type-II parametric down conversion
  source in a periodically poled KTP waveguide at telecom wavelengths,
  with almost perfect photon-number correlations.  We used a
  photon-number-resolving time-multiplexed detector based on a
  fiber-optical setup and a pair of avalanche photodiodes.  By
  resorting to a germane data-pattern tomography, we assess the
  properties of the nonclassical light states states with
  unprecedented precision.
\end{abstract}


\maketitle

\textit{Introduction.---}
Nonclassical states of light constitute an invaluable resource for
deploying quantum-enhanced technologies as diverse as cryptography,
computing, and metrology, to cite only some of the many relevant
examples. Certifying signatures of nonclassicality generally requires
inferring either the photon-number distribution or a quasiprobability
distribution indirectly from a set of measurements.  Even though the
latter approach is well established~\cite{Lvovsky:2009fk} (it involves
homodyne detection followed by an appropriate reconstruction scheme),
photon counting seems a more natural choice in this discrete-variable
scenario, in which photons are used as flying qubits.

However, capitalizing on photon counting places stringent demands on
detector performance, quantified in terms of, e.g., spectral range,
efficiency, dead time, dark-count rates, and timing jitter. This is
currently driving considerable improvements in single-photon
detectors~\cite{Hadfield:2009qr,Buller:2010cl,Eisaman:2011tg,
  Sprengers:2011dk,Natarajan:2012bf,Pernice:2012kb,Calkins:2013sf,
  Migdall:2013ef,Natarajan:2013bh}; in particular, the photon-number
resolving (PNR) capability is nowadays required in most advanced
protocols.

Several strategies have been proposed thus far for PNR detectors.
Single-photon avalanche diodes (SPADs) have become the prevailing
option for PNR applications. Si-based SPADs constitute a relatively
mature technology with several efficient devices commercially
available, but they are only suitable for use at visible and near
infrared wavelengths. For experiments at technologically-important
telecom wavelengths, the main contending technologies are InGaAs
SPADs, which are plagued by high dark-count rates and long dead times,
thereby making gating essential.

A proposal to employ a time-multiplexed detection (TMD) based on
SPAD has been put recently forward~\cite{Achilles:2003cs,
  Rehacek:2003bh,Fitch:2003uq}.  These TMDs work also for pulsed
light, and the photon-number distribution of a quantum state can be
retrieved by inverting the measured photon statistics. Experimental
applications, demonstrating a reliable loss calibration, and the TMD
suitability for detecting multimode statistics and nonclassicality,
have already been accomplished~\cite{Achilles:2006fk,Avenhaus:2008fk,
Worsley:2009qr,Avenhaus:2010ff,Laiho:2010kl,Bartley:2013rz}.

The effective implementation of these advanced schemes relies on a
complete and accurate knowledge of the detector, an issue that has
lately started to attract a good deal of attention~\cite{Luis:1999yg,
  Fiurasek:2001dn,DAriano:2004oe,Lundeen:2009sf,Amri:2011fk,
  Zhang:2012fu,Brida:2012mz}. The idea behind is to employ the outcome
statistics in response to a set of complete certified input states.
 
However, as shown in Ref.~\cite{Rehacek:2010fk}, if the measurement
itself is of no interest, the costly detector calibration can be
bypassed by using a direct fitting of data in terms of detector
responses to input probes.  Thus, state estimation is done without any
prior knowledge of the measurement, avoiding unnecessary wasting of
resources on evaluating the parameters of the
setup~\cite{Mogilevtsev:2013kl,Cooper:2013fq}.

In this Letter, we present a thorough application of this novel
data-pattern tomography to TMDs. In this way, we provide a full
account of the nonclassical properties of quantum states.

\begin{figure}
  \begin{center}
    \includegraphics[width=0.99\linewidth]{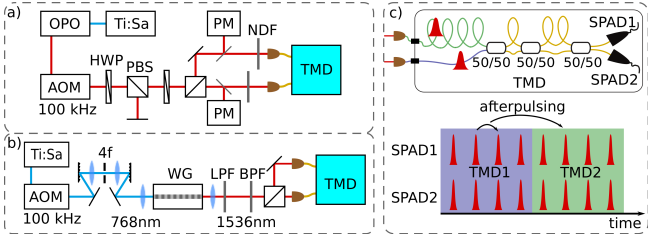}
    \caption{(Color online) Left panel: (a) Probe-state generation.
      Pulsed light at telecom wavelengths is generated in a
      Ti:Sapphire pumped optical parametric oscillator (OPO). The
      repetition rate of the reference is lowered by an
      acousto-optical modulator (AOM). Motorized half-wave plates
      (HWP) followed by polarizing beam splitters (PBS) are used to
      vary the attenuation. The light power is measured by power
      meters (PM), further attenuated by neutral density filters
      (NDF), and coupled into the single-mode fibers of the
      time-multiplexing detector (TMD).  (b) Nonclassical state
      generation. The pump light is spectrally tailored by a $4f$
      system, coupled into the KTP waveguide (WG) and blocked by a
      long pass filter. The PDC photons pass a bandpass filter (BPF),
      are separated by a PBS and coupled into the TMD. Right panel:
      TMD. Two input pulses are distributed into 16 bins to obtain
      information about the photon number in each input as well as
      photon number correlations between the beams.}
    \label{fig:TMD}
  \end{center}
\end{figure}

\textit{Experimental setup.---}
The states in our experiment are generated by type II parametric down
conversion (PDC) inside a periodically poled KTP waveguide. The PDC
source produces decorrelated signal and idler states with a purity for
heralded states above $80\%$ and high coupling efficiency into
single-mode fibers. The setup is the same as the one described in
detail in Ref.~\cite{Harder:2013tx} and  sketched in
Fig.~\ref{fig:TMD}.

Twin beams created in PDC are archetypal example of highly
correlated quantum states. Sub- and super-Poissonian photon
statistics~\cite{Riedmatten:2004dz},
antibunching~\cite{Koashi:1993fu}, and quantum correlated quadrature
amplitudes~\cite{Rarity:1992fe} have been demonstrated.

Our TMD is also schematized in Fig.~\ref{fig:TMD}. Two incoming pulses
are split into 16 temporal bins and impinge onto SPADs. Counting the
clicks allows us to estimate photon numbers and photon-number
correlations between the two input ports. Since we work at telecom
wavelengths, we use InGaAs SPADs (Id Quantique id201 at a repetition
rate of 1 {MHz} with a gate width of about 2.5 {ns}). As briefly
mentioned before, InGaAs SPADs SPADs are the simplest and most
cost-efficient detectors available at telecom wavelengths. However,
they have some disadvantages: the detection efficiencies are below
$25\%$ and afterpulsing is present with a few percent
probability~\cite{Kang:2003kh}. Consequently, the conventional TMD
model~\cite{Achilles:2003cs}, which only takes into account the
probabilistic splitting and overall losses, appears to be
inadequate. A more sophisticated technique is required to recover
photon statistics from the measured click frequencies; this is where
data-pattern tomography comes into play.

The state is specified by the two-mode photon-number distribution
$P_{mn}$, where the first (second) index refers to the signal
(idler) mode. We also denote by $p_{\alpha\beta}$ the probability of
simultaneous signal ($\alpha$) and idler ($\beta$) detection.
Detections are thus described by 8-digit binary numbers, where $0$/$1$
values mean click/no click in the corresponding time bin. For example, 
$\beta=00000011$ denotes a simultaneous detection in the first two
idler time bins. This gives $2^8=256$ distinct single-mode events and
$2^{16}=65536$ two-mode  events to reckon with.

We adopt a linear model of the TMD detection, so that
\begin{equation}  
p_{\alpha\beta} = \sum_{m=0}^{d-1}\sum_{n=0}^{d-1} C_{\alpha\beta,mn}
\, P_{mn} \, ,
\end{equation}
where $d$ is the cutoff dimension required to accommodate the relevant
parts of the signal and the idler and  the measurement matrix $C$
provides a complete description of the TMD, including losses, detector
efficiencies and afterpulsing effects. 

In a real experiment, we acquire the relative frequencies
$f_{\alpha\beta}$ after $N$ random samples drawn from the multinomial
distribution parametrized by $p_{\alpha\beta}$.  Due to
afterpulsing, it is not possible to factorize the detection matrix in
signal and idler parts. 

We also consider single-mode and heralded events; the former (latter)
are simply marginal (conditional) probabilities of $p_{\alpha\beta}$.
For these single-mode events, we look at the total number of clicks
(either in the signal or the idler), without paying attention to the
particular ordering of time bins.  For example, for the signal-mode
reconstruction, such reduction is readily done by summing data and
patterns marginals $f_\alpha=\sum_\beta f_{\alpha\beta}$ over the
$8!/[(8-k)! k!]$ different permutations of $\alpha$ with the same
number $k$ of nonzero binary digits.

\textit{Fitting data patterns.---}
From the measured data $f_{\alpha\beta}$ we have to determine the
state $P_{mn}$. The standard detector tomography would proceed in
two steps: first, a detector estimation, where the measurement matrix
$C_{\alpha\beta,mn}\ge 0$ is inferred from a set of calibration
states. Afterwards, the state $P_{mn}\ge 0$ is reconstructed from 
the previously obtained  detector matrix.  However,
this is not completely satisfactory: the details of the TMD are not
of interest and, besides, the detector estimation is exceedingly
costly, scaling as $d^4$, which makes the method impractical, even for
moderate values of this cutoff $d$.

The alternative data-pattern approach, we adopt here, expresses
$P_{mn}$ as a mixture
\begin{equation}
  \label{decomp}
  P_{mn} = \sum_{\xi=1}^M x_{\xi} \, P^{(\xi)}_{mn} =  
  \sum_{\xi=1}^M x_{\xi} \, P_m^{(\xi)} \;  P_n^{(\xi)} \, ,
\end{equation}
of $M$ linearly independent (generally, nonorthogonal) two-mode
coherent probes $\{ P^{(\xi)}_{mn} \}$, with positive and negative
weights $\{x_{\xi} \}$. This discrete representation can be thought of
as a kind of generalization of the $P$-representation and can be
sufficiently accurate depending on the number of terms in the sum.

The responses $f_{\alpha\beta}^{(\xi)}$ of the TMD to these coherent
probes are called patterns. Then, by linearity, the data (i.e., the
TMD response $f_{\alpha\beta}$ to an unknown state $P_{mn}$)
can be modeled in terms of patterns as
\begin{equation}
  \label{fit}
  f_{\alpha\beta} \simeq  \sum_{\xi=1}^{M}  x_{\xi} \,  
  f_{\alpha\beta}^{(\xi)} \, . 
\end{equation}
Hence, once the patterns and data are measured, the coefficients
$x_{\xi}$ can be inferred from Eq.~\eqref{fit} and the state
reconstructed according to \eqref{decomp}. To this end, a suitable
convex measure of the distance between the left- and right-hand side
of Eq.~\eqref{fit} has to be minimized, subject to the physical
constraints $P_{mn}\ge 0$ and $\sum_{mn}P_{mn}=1$: this is a quadratic
program than can be efficiently solved~\cite{Motka:2014il}.

Notice that in the data-pattern tomography, the
number of parameters $M-1$ is independent of the probe cutoff
dimension $d$. Also, if needed, a partial tomography of the unknown
state can be performed by using only a small part of the patterns or 
any linear function of them (such as the value of the Wigner function
at the origin)  for the data fitting in Eq.~\eqref{fit}.  

To create the probe states we use pulsed coherent light attenuated
at the single-photon level. The
power of the reference beam is changed by two motorized half-wave
plates followed by polarizing beam splitters. We calibrate all the
neutral density filters separately and measure fiber-coupling
losses. From these values and the measured reference power, we 
calculate the power inside the fibers of the TMD. Due to the high
degree of attenuation (of the order of $10^{-9}$), small calibration
errors (of order of a few percent) cannot be avoided. However, this
affects the total losses, but not the shape of the photon statistics.

\textit{Results.---} 
We take into consideration a fixed number of patterns with amplitudes
below a given threshold $\alpha_\text{max}\approx 2$.  This threshold
is important because of the afterpulsing, which seems to be more
pronounced for stronger states.  The reconstruction is repeated 100
times with randomly chosen probe subsets of size $M<235$ and averaged
over those repetitions. In this way the redundancy in the data is
propagated into the final estimate. 

The variation within the set of reconstructions is used to estimate
the associated errors, much in the spirit of nonparametric
bootstrap~\cite{Davison:1997ri}.  In the experiment, $N_\xi=4.2\times
10^6$ events were registered for each coherent probe and PDC state.
For low-intensity PDC states, the data were averaged over five
repeated data acquisitions, making a total of $N_\text{PDC} = 21
\times 10^6$ events. With these numbers, the statistical noise is
insignificant (except, perhaps, for heralded detections) and the
reconstruction accuracy is governed by systematic errors and
afterpulsing effects.

To check the performance for different parameter sets we first
performed a cross-validation~\cite{Mogilevtsev:2013hb}, to verify
whether the estimated state is consistent with the observed data
sample.  To this end, we checked the quality of the reconstruction
with random sets of coherent states discarded from the probes, but
with the same amplitude threshold. We have resorted to different
measures of errors; for all of them we conclude that the accuracy is
insensitive to the dimension $d$, while the errors get larger for
stronger probes. Typical fidelities around $99\%$ are attained, which
amounts to errors of a few percent for the reconstructed elements of
$P_{mn}$, which outperforms the standard detector tomography.

More probes give somewhat better results, but small sets of probes can
be surprisingly good.  This is due to the small variation across those
patterns characterized by a small  number of principal components in
the singular value decomposition. 

\begin{figure}
  \centerline{
    \includegraphics[width=0.49\linewidth]{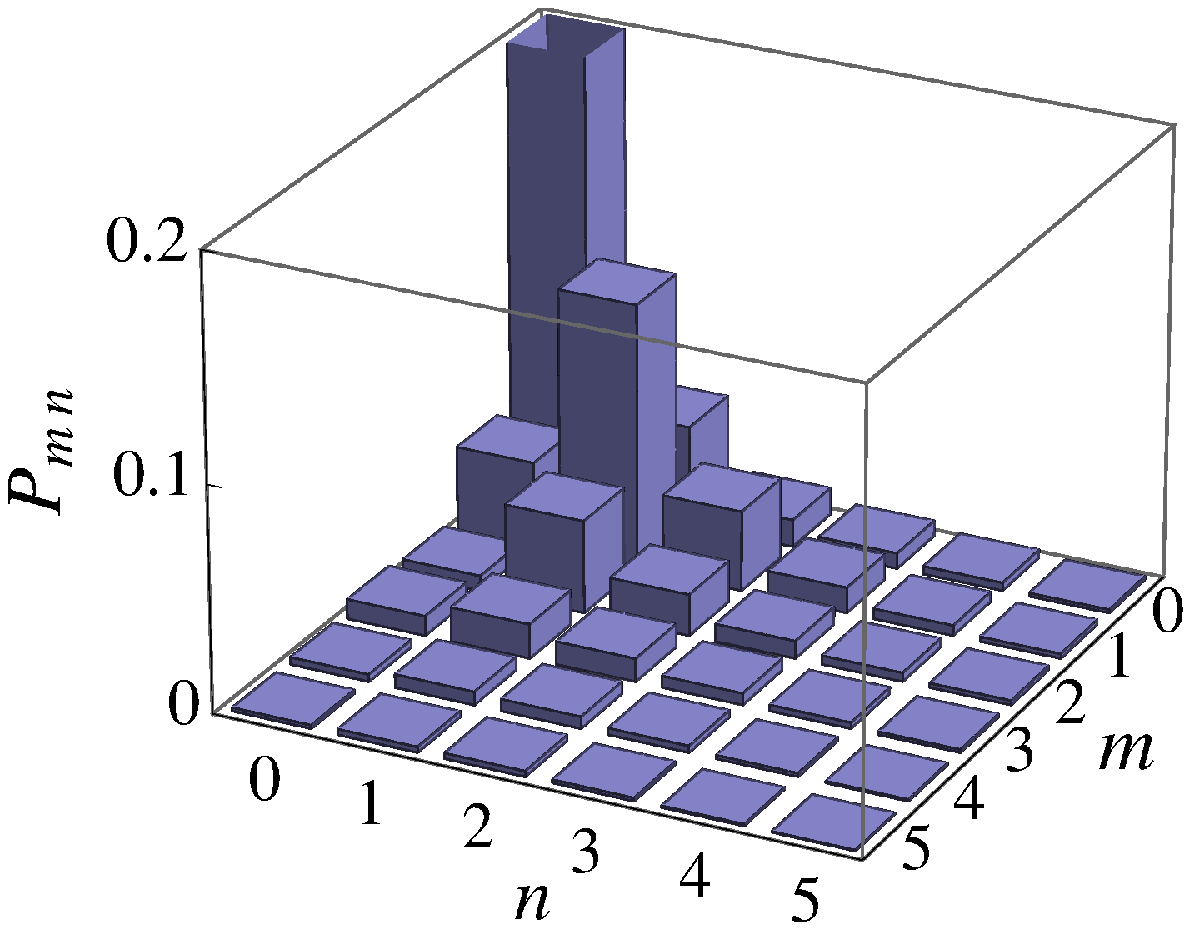}
    \hfill
    \includegraphics[width=0.49\linewidth]{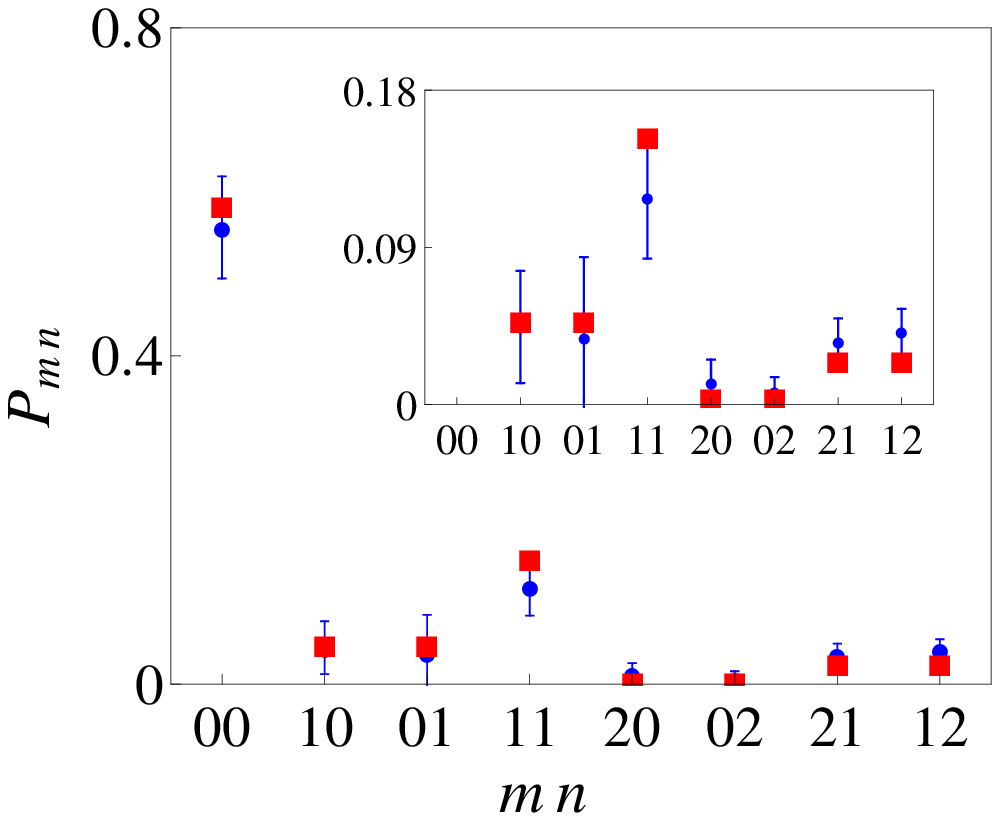}}
  \caption{(Color online) Two-mode reconstruction of the PDC$_{2}$
    state.  Left panel: Complete photon-number distribution, with
    $M=50$ probes.  Right: Comparison with the theoretical model. Circles
    with error bars represent the experimental results, whereas theory
    is indicated by squares.}
  \label{fig:2D}
\end{figure}

To compare with the theory, we assume the PDC distribution
\begin{equation}
  \label{pdc}
  \mathcal{P}_{mn}=\frac{\langle n\rangle^n} 
  {(1+\langle n\rangle)^{1+n}} \delta_{mn} \, ,
\end{equation}
together a finite detection efficiency  $\eta$ which is
taken into account by a Bernoulli distribution:
\begin{equation}
  \label{coupling}
  P_{mn} = \sum_{k =m}^{\infty} \sum_{\ell =n}^{\infty}
  \left( 
    \begin{array}{c}
      k \\ m
    \end{array}
  \right)
  \left (
    \begin{array}{c}
      \ell \\ n
    \end{array}
  \right) 
  \eta^{n+m} (1-\eta)^{k+\ell -n-m}
  \mathcal{P}_{k\ell} \, .
\end{equation} 
From the zero-detection probabilities of coherent probes with known
amplitudes, the quantum efficiency of detectors was estimated to be
$0.22 \pm 0.01$ and the coupling efficiency $75\%$. This, in turn,
enables to calculate the mean photon numbers of the generated PDC
states.  Three PDC, denoted PDC$_1$, PDC$_2$ and PDC$_3$, were
generated, with $\langle n_{1} \rangle=0.11$, $\langle n_{2} \rangle=0.76$,
and $\langle n_{3} \rangle=1.34$, respectively. These numbers were used
to predict the two-mode statistics through Eqs.~\eqref{pdc} and
\eqref{coupling}.

In Fig.~\ref{fig:2D} we plot typical results of two-mode TMD
measurements for PDC$_{2}$. Strong signal-Idler correlations are
observed  and the agreement with the theory is pretty good. Similar
results are found for other intensities.

  \begin{figure}
    \begin{center}
      \includegraphics[width=0.49\linewidth]{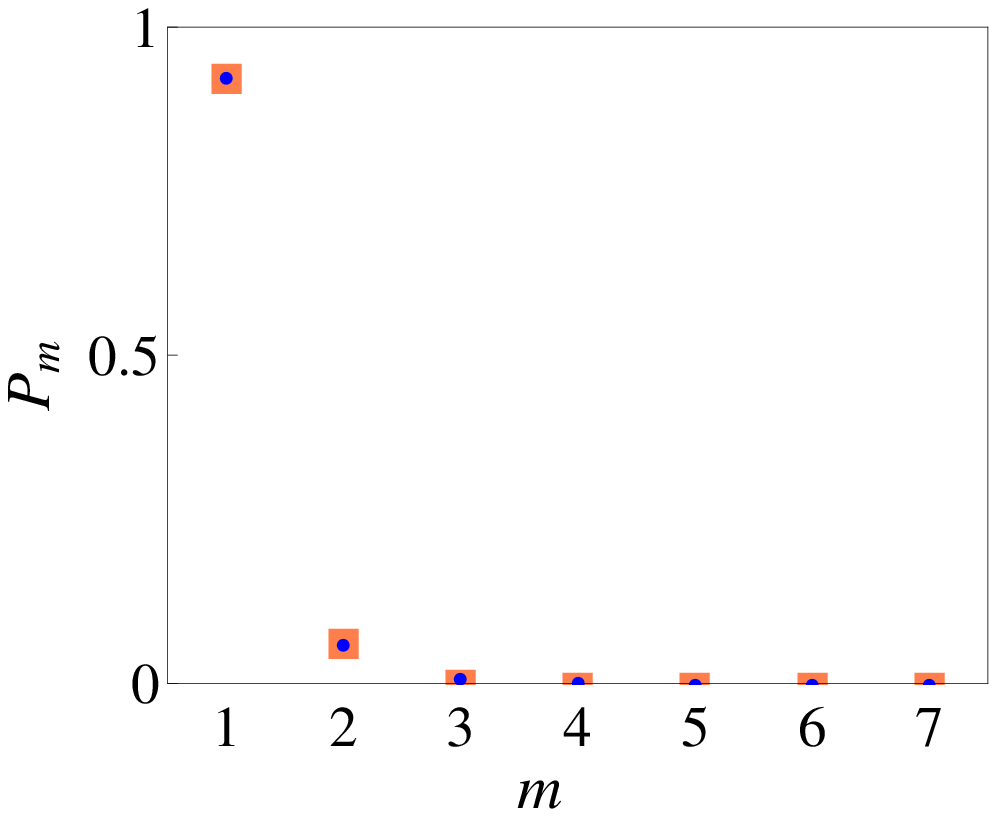} \hfill
      \includegraphics[width=0.49\linewidth]{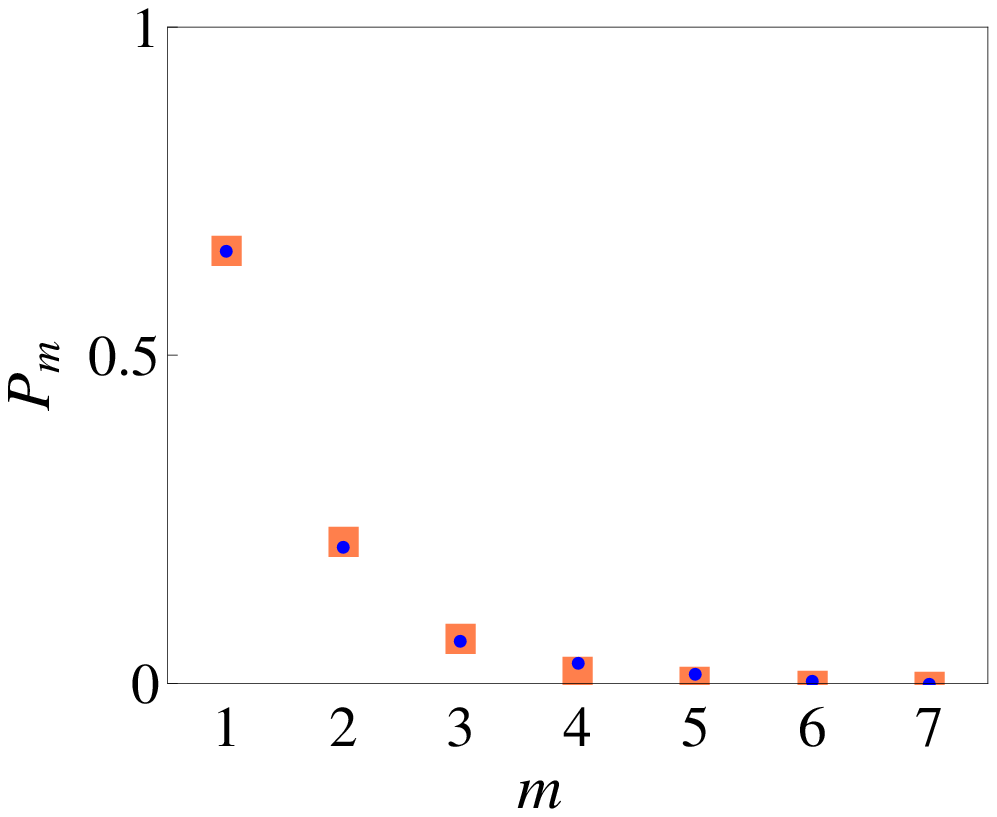}
      \caption{(Color online) Single-mode signal reconstructions
        (dots) of PDC$_1$ (left) and PDC$_2$ (right), both with $M=30$
        probes. Best fits to Bose-Einstain distributions (squares)
        are also shown. The reconstruction errors are almost
        negligible and cannot be appreciated.}
      \label{fig:PBose}
    \end{center}
  \end{figure}

  In Fig.~\ref{fig:PBose} we show the reconstructions of the signal
  states for two different pump intensities. Best fits to
  Bose-Einstein distributions are almost indistinguishable from the
  experimental results.

  Heralded states are created by having the idler state conditioned
  on single or double detection in the signal of the PDC output.
  By double detection we mean here a click at detector $A$ accompanied
  by a simultaneous click at detector $B$. Double detections at any
  single detector are discarded to avoid doubles caused by 
  afterpulsing.

Heralded single- and especially two-photon states are difficult to
reconstruct, since we are picking out quite a small subset of all the
detection events.  Besides, afterpulsing creates artificial signal-idler
correlations, whose strength depends on the distance of the signal
detection from the first idler time bin. All in all, this leads to larger
reconstruction errors compared to single-mode states. 

\begin{figure}
  \begin{center}
    \noindent
    \includegraphics[width=0.49\linewidth]{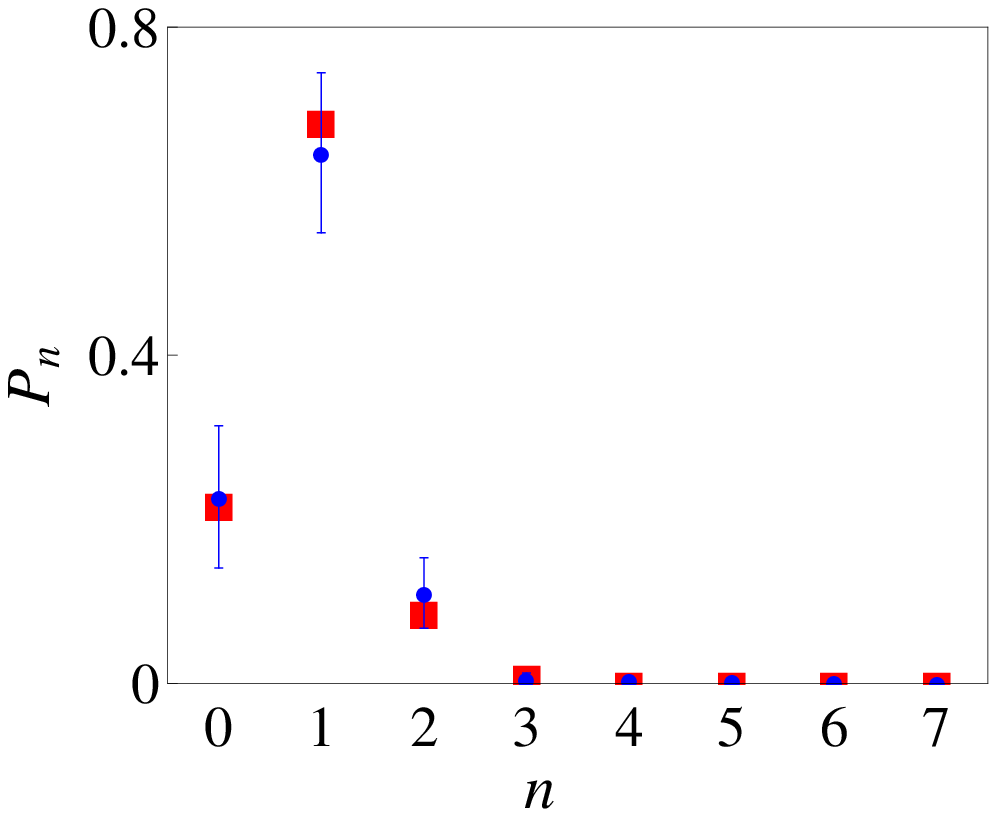}%
    \hfill
    \includegraphics[width=0.49\linewidth]{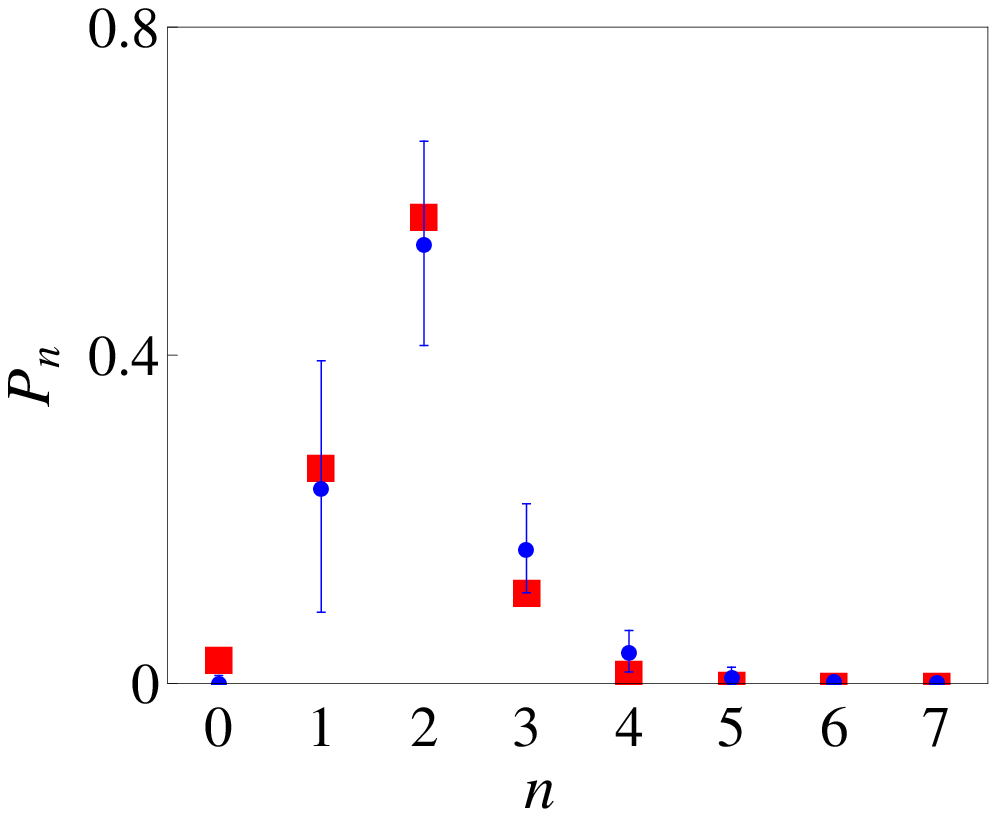}\\
    \includegraphics[width=0.49\linewidth]{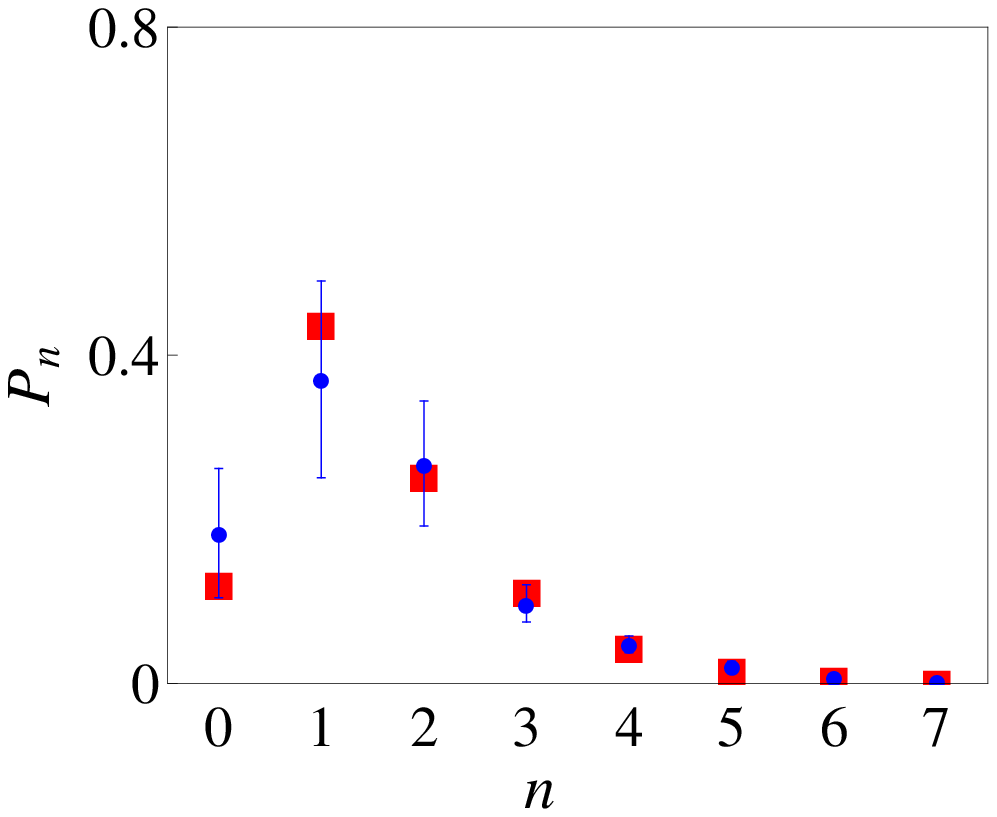}%
    \hfill
    \includegraphics[width=0.49\linewidth]{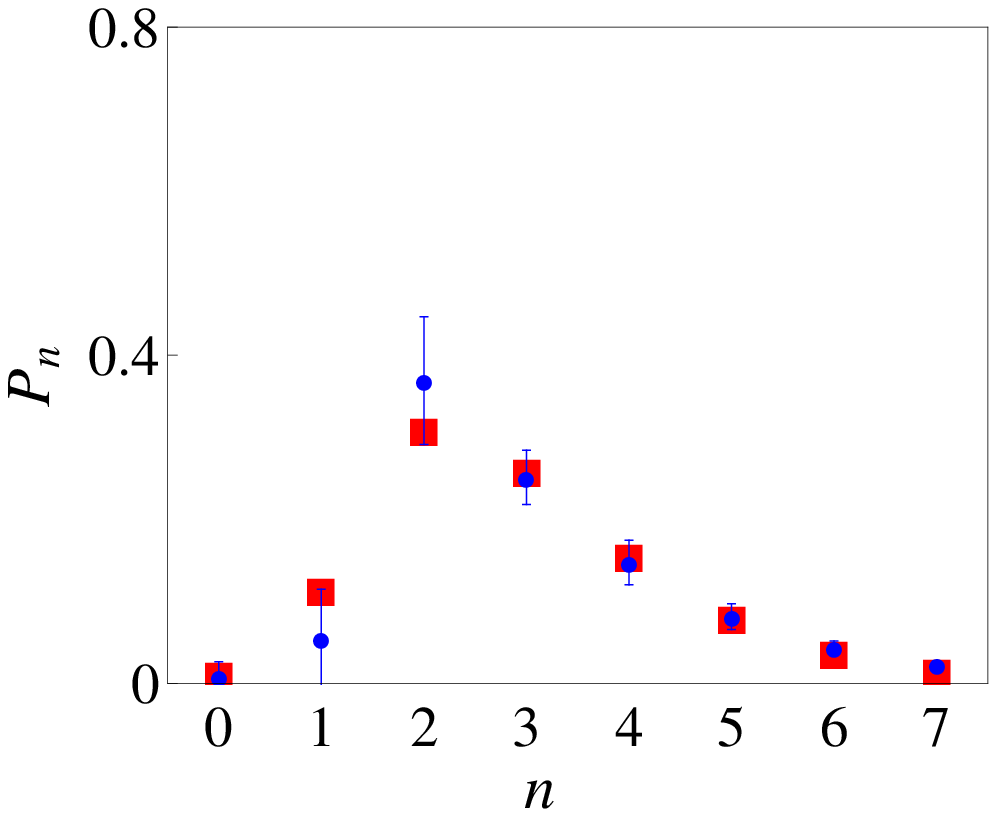}
    \caption{(Color online) Reconstructed single-photon (left) and
      two-photon (right) heralded idler states generated from PDC$_1$
      (top) and PDC$_2$ (bottom), with $M=80$ probes. Squares denote
      again the corresponding theoretical predictions.}
    \label{fig:heralded}
  \end{center}
\end{figure}

Reconstructed single-and two-photon heralded idler states from two
different PDC states are shown in Fig.~\ref{fig:heralded}. To get
theoretical predictions, we again assume an inefficient coupling
($0.75$) of the PDC state and calculate the post-measurement idler
state $P_{i}$ from the pre-measurement $P$ as follows
\begin{equation}
  P_{i} =\frac{ \Tr_{s} ( E \,  P \, E^{\dagger})} 
  {\Tr_{s,i} (E  \, P  \,  E^{\dagger}) },
\end{equation}
where $\op{E}^\dag \op{E}$ is the POVM element describing the
single/double detection in the signal mode and $\Tr_{s,i}$ indicates
trace over the signal/idler. All states and POVM elements are diagonal
here.

Best estimates of Wigner function at the origin for the single-photon
heralded states are $W(0)=- 0.72\pm 0.06$ (PDC$_1$) and $W(0)=-0.30\pm
0.09$ (PDC$_2$).  This agrees with the calculated values $W(0)=-0.77$
(PDC$_1$) and $W(0)=-0.29$ (PDC$_2$), respectively and confirms the
noclassicality of these states. With more intense PDC inputs, single
detection in the signal tends to leave a mixture of Fock states in the
idler. This explains why the nonclassicality of heralded states
decreases with increasing pump intensity.

\begin{figure}[t]
  \begin{center}
    \includegraphics[width=0.49\linewidth]{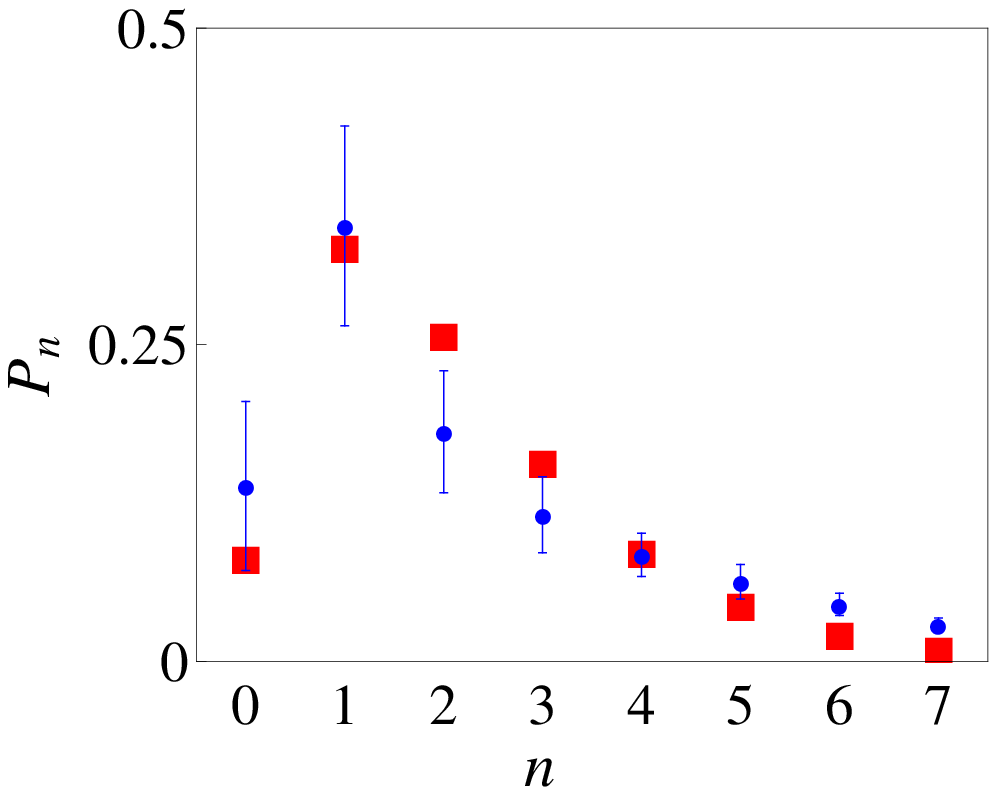}
    \hfill
    \includegraphics[width=0.49\linewidth]{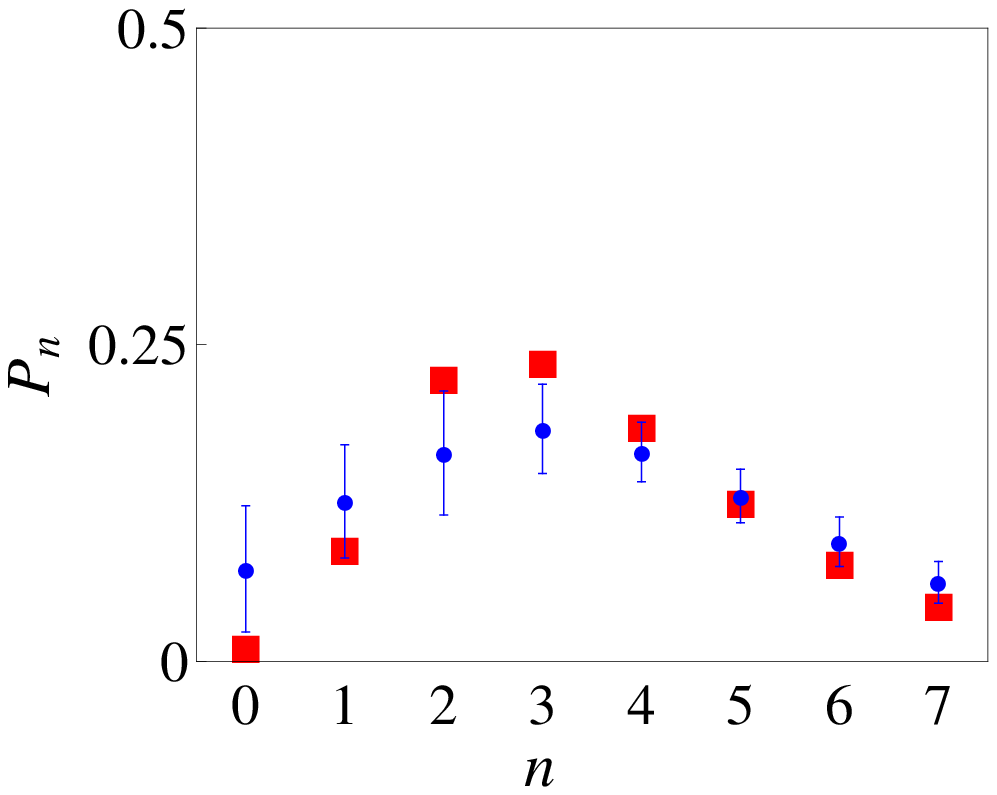}
    \caption{(Color online) Heralded single-photon (left) and
      double-photon (right) Idler states as predicted from the
      reconstructed two-mode photon-number distributions of PDC$_3$,
      with $M=80$ probes.}
    \label{fig:simulated}
  \end{center}
\end{figure}

Finally, we  simulated heralded states as post-measurement states based
on the results of full two-mode tomography.  To this end, we performed
100 two-mode reconstructions for each measured PDC state. The idler
post-measurement state is calculated based on a thought single or
double signal detection. The statistics of the resulting ensemble of
heralded states is shown in Fig.~\ref{fig:simulated}, where we compare
this statistics with the theoretical predictions.

These predictions based on the full two-mode reconstructions are less
accurate than the single-mode heralded ones.  The latter is more
direct.  In heralded detections, what helps is that the dimension of
the search space is reduced and the dominating vacuum or even
single-photon terms are eliminated, which improves the accuracy. In
addition, in the data-pattern approach we use heralded coherent probes
i.e. do the same data selection as for the PDC data. In this way, one
somehow eliminates the artificial correlations created by the
afterpulsing.  Nevertheless, it is nice to see that the accord between
single- and two-mode measurements is actually pretty good. One can
also notice that the two-mode predictions improve with increasing
intensity, as one could expect.  More intense PDC states have larger
higher-order $P_{mn}$ components, which are easier to extract.

\textit{Concluding remarks.---} 
In summary, we have exploited a PDC source of quantum states at
telecom wavelengths with remarkable properties in terms of brightness,
purity and symmetry. To put forward the nonclassical issues of the
generated states, we have employed TMDs together with the method of
data-pattern tomography. The experimental calibration shown here goes
beyond any quantum detector tomography previously demonstrated.  Our
approach is easily adapted to a variety of measurement devices and the
experimental implementation presented here shows its viability for
complex detectors.

J.~R., Z.~H, L.~M., and B.~S. thank the financial assistance of the
Technology Agency of the Czech Republic (Grant TE01020229) and the IGA
Project of the Palack\'y University (Grant PRF 2013 019). L.~L.~S.~S acknowledges the support from the Spanish MINECO (Grant
FIS2011-26786).

%

\end{document}